\documentclass[11pt]{article}
\usepackage{moriond,epsfig,subfigure}

\def\be{\begin{equation}}
\def\ee{\end{equation}}
\def\bea{\begin{eqnarray}}
\def\eea{\end{eqnarray}}

\begin{document}
\vspace*{4cm}
\title{Present and possible future implications for mSUGRA of the non-discovery of SUSY at the LHC}

\author{P.~Bechtle\,\footnotemark[1], K.~Desch\,\footnotemark[2], H.~Dreiner\,\footnotemark[2], M.~Kr\"amer\,\footnotemark[3], B.~O'Leary\,\footnotemark[4], C.~Robens\,\footnotemark[3], B.~Sarrazin\,\footnotemark[1], P.~Wienemann\,\footnotemark[2]\\[3ex]}
\footnotetext[1]{Deutsches Elektronen-Synchrotron DESY, Notkestra{\ss}e 85, D-22607 Hamburg, Germany}
\footnotetext[2]{Physikalisches Institut der Universit\"at Bonn, Nussallee 12, D-53115 Bonn, Germany}
\footnotetext[3]{Institute for Theoretical Particle Physics and Cosmology, RWTH Aachen, D-52056 Aachen, Germany}
\footnotetext[4]{Institute for Theoretical Physics and Astrophysics, W\"urzburg University, Am Hubland, D-97074 W\"urzburg, Germany}
\setcounter{footnote}{4}


\maketitle


\abstracts{ Both ATLAS and CMS have published results of SUSY searches
  putting limits on SUSY parameters and masses. A non-discovery of
  SUSY in the next two years would push these limits further. On the
  other hand, precision data of low energy measurements and the dark
  matter relic density favor a light scale of supersymmetry. Therefore
  we investigate if supersymmetry -- more specifically the highly
  constraint model mSUGRA -- does at all agree with precision data and
  LHC exclusions at the same time, and whether the first two years of
  LHC will be capable of excluding models of supersymmetry. We
  consider the current non observation of supersymmetry with 35~pb$^{-1}$
  as well as the possible non observation with 1, 2 and 7~fb$^{-1}$ in a
  global fit using the framework Fittino.  }

\section{Introduction}\label{sec:intro}


Supersymmetry\,\cite{Nilles:1983ge} (SUSY) provides an elegant and
renormalizable solution to several current problems of the Standard
Model (SM) of elementary particles: Provided its parameters are chosen
appropriately, it can explain electroweak symmetry breaking, solve the
hierarchy problem of the Higgs sector of the SM, and provide the
correct amount and structure of Dark Matter (DM) in the universe,
together with agreement of its predictions with precision measurements
at various experiments. However, all these are typically only
fulfilled simultaneously for very specific parameter settings and
breaking assumptions. 

Many previous studies of the available data before the LHC\,\cite{deAustri:2006pe,Allanach:2007qk,Lafaye:2007vs,Buchmueller:2008qe,Bechtle:2009ty,Buchmueller:2009fn} era indicate that a mass scale
of the SUSY particles below around 1.6~GeV is required at the
2~$\sigma$ level to bring a highly constrained model such as
mSUGRA/CMSSM in agreement with all precision results. Strong
constraints are placed on details of the mass spectrum and the
couplings, e.g. corresponding to a co-annihilation process to control
the DM content (see e.g.\,\cite{Bechtle:2009ty}).

Since SUSY is already highly constrained before including the present
non-observation of new physics at LHC in the fit, it is a highly
non-trivial question whether SUSY can be brought in agreement with
both the LHC limits and the precision data, even though the upper
mass bound on the colored particles from the previous fits is
considerably higher than the present LHC limits (see
e.g.\,\cite{daCosta:2011qk,Khachatryan:2011tk}), at around 800~GeV,
since the precision data also put constraints on details of the
model, as described above.

The analysis presented here\,\cite{Bechtle:2011dm} is using the mSUGRA
model to answer the question of the level of agreement for the
following reason: if this highly constrained (but well-understood)
model is in agreement with the data, then more general SUSY models
will be in agreement, too. If not, other breaking scenarios and
generalizations of mSUGRA with weaker high-scale assumptions will have
to be tested.


At LHC, SUSY can be searched in different channels asking for varying
numbers of hard jets, leptons and amounts of missing transverse
energy. The strongest constraints are currently stemming from very
inclusive analyses asking only for jets and leptons. In addition, such
analyses have the advantage that their results do depend on only few
mSUGRA parameters. Therefore, a study for inclusive searches at
ATLAS\,\cite{Atlas-pub} has been modeled as a prediction for the
actual results of the experiments, since the fits presented here were
performed in parallel to the presently public searches by ATLAS and
CMS. For other recent contributions in the same field, see
e.g. Ref.\cite{others}

\section{Model, Inputs to the Fit and Statistics}\label{sec:model}

The mSUGRA model used in the fit are evaluated using a Markov Chain
Monte Carlo technique. The theoretical predictions are calculated
using SPheno\,\cite{Porod:2003um} for the RGE running and the spectrum
calculation, and programs compiled in the mastercode package for the
prediction of the low energy precision observables and the Higgs boson
masses, most notably FeynHiggs, micrOmegas and
SuperISO\,\cite{Buchmueller:2009fn}. SoftSUSY\,\cite{Allanach:2001kg}
is used for cross-checks.

\subsection{Observables from the pre-LHC era}\label{sec:LEobs}

Following the Fittino\,\cite{Bechtle:2004pc} analysis in Ref.\,\cite{Bechtle:2009ty}, the
following set of low-energy observables and existing collider limits
is used: \textit{(i)} rare decays of B- and K-mesons; \textit{(ii)}
the anomalous magnetic moment of the muon, $a_\mu$; \textit{(iii)}
electroweak precision measurements from LEP, SLC and the Tevatron and
the Higgs boson mass limit from LEP; and \textit{(iv)}~the relic
density of cold dark matter in the universe, $\Omega_\chi$. In
contrast to Ref.\,\cite{Bechtle:2009ty}, we employ the program
HiggsBounds\,\cite{Bechtle:2008jh} and not a rigid Higgs mass limit. We
refer to Ref.\,\cite{Bechtle:2009ty} for a detailed discussion of the
low-energy inputs and the collider limits.

\subsection{Modeling the ATLAS analysis}\label{sec:modelling}

The most sensitive and at the same time rather model independent
search channel for $R$-parity conserving SUSY at the LHC relies on
jets and missing transverse energy $E_T^{\rm miss}$ for the
selection. From the analyses presented in the ATLAS MC
study\,\cite{Atlas-pub}, we consider the search channel with four
jets, zero leptons and $E_T^{\rm miss}$.  This channel drives the
sensitivity, in particular for large $M_{1/2}$. For a detailed
description of the selection cuts applied see Ref.\,\cite{Bechtle:2011dm}. 
As a final discriminating variable the effective mass is used. It is
defined as the scalar sum of the transverse momenta of all main
objects, \textit{i.e.}\
\begin{equation}
M_{\mathrm{eff}}=\sum_{i=1}^{N_{\rm jets}=4} p_T^{\mathrm{jet},i}
+E_T^{\mathrm{miss}}.
\end{equation}
The SM background processes have been described in detail in
Ref.\,\cite{Atlas-pub}. We use the background shape of
$M_{\mathrm{eff}}$ from the ATLAS analysis directly in our fit. A
systematic uncertainty of 20\,\%, derived from Ref.\,\cite{Atlas-pub}
has been used on the background, which is also in rough agreement with
the present results based on data\,\cite{daCosta:2011qk}.  The signal
cross section is dominated by squark and gluino pair production, $pp
\to \tilde q \tilde q^*, \tilde q \tilde q,\tilde q \tilde g$ and
$\tilde g\tilde g$, but all SUSY pair production processes are
included in our numerical analysis. We use Herwig{\footnotesize
  ++}\,\cite{Bahr:2008pv} in combination with the parametrized fast
detector simulation Delphes\,\cite{Ovyn:2009tx} to obtain the detector
response and, in particular, the shape of the $M_{\rm eff}$
distribution for a given point in the supersymmetric parameter space.
The simulation has been carefully modified to match the published
measured resolutions and efficiencies of the ATLAS experiment, and the
resulting $M_{\rm eff}$ has been compared to the public spectra at an
mSUGRA benchmark point. The signal estimate is normalized to the
NLO+NLL QCD prediction for the inclusive squark and gluino cross
sections~\cite{NLLrefs}.

\begin{figure}[t]
  \centering
  \subfigure[ ]{
    \includegraphics[width=0.45\textwidth]{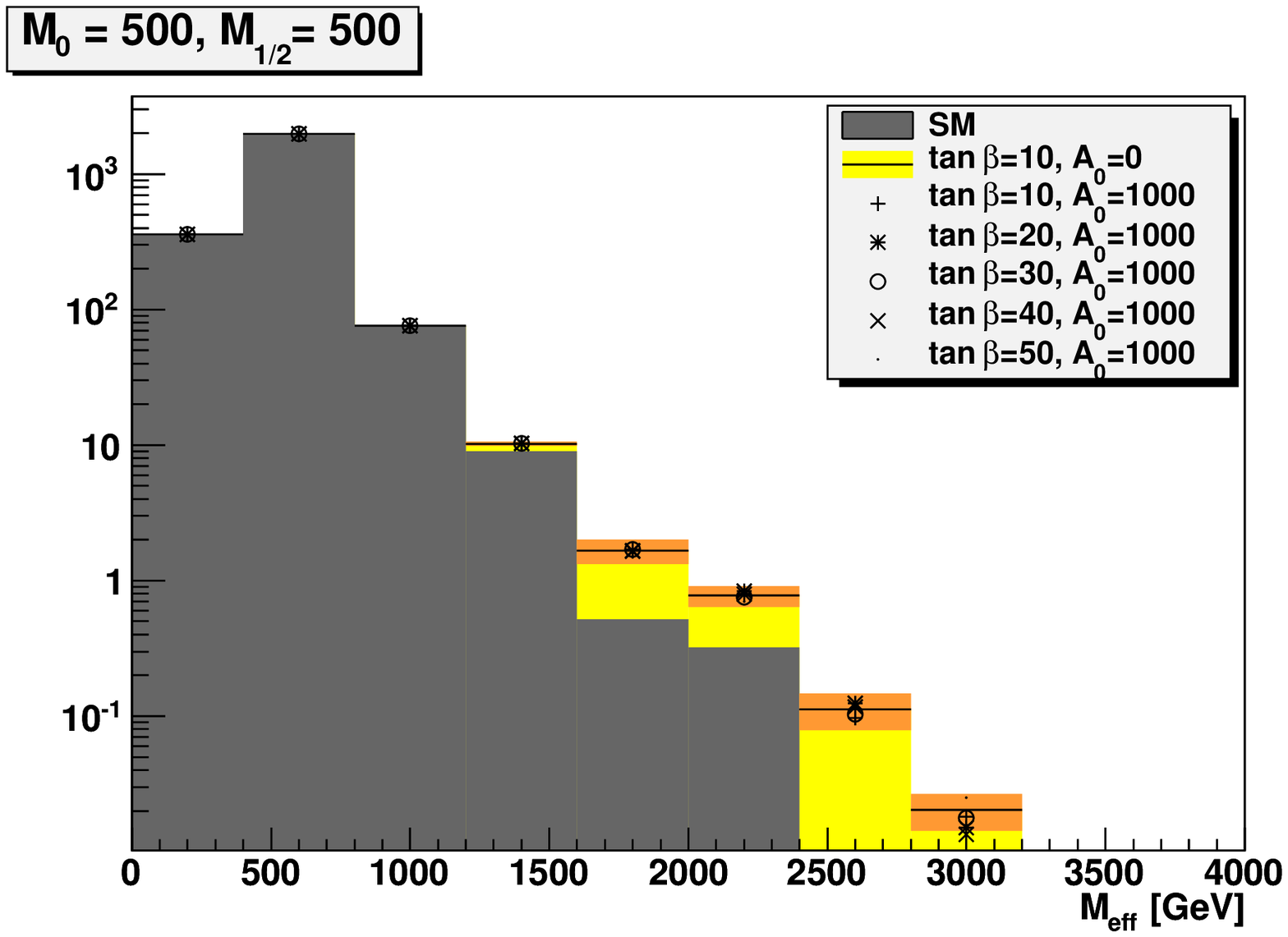}
    \label{fig:Syst_subfig1}
  }
  \subfigure[ ]{
    \includegraphics[width=0.45\textwidth]{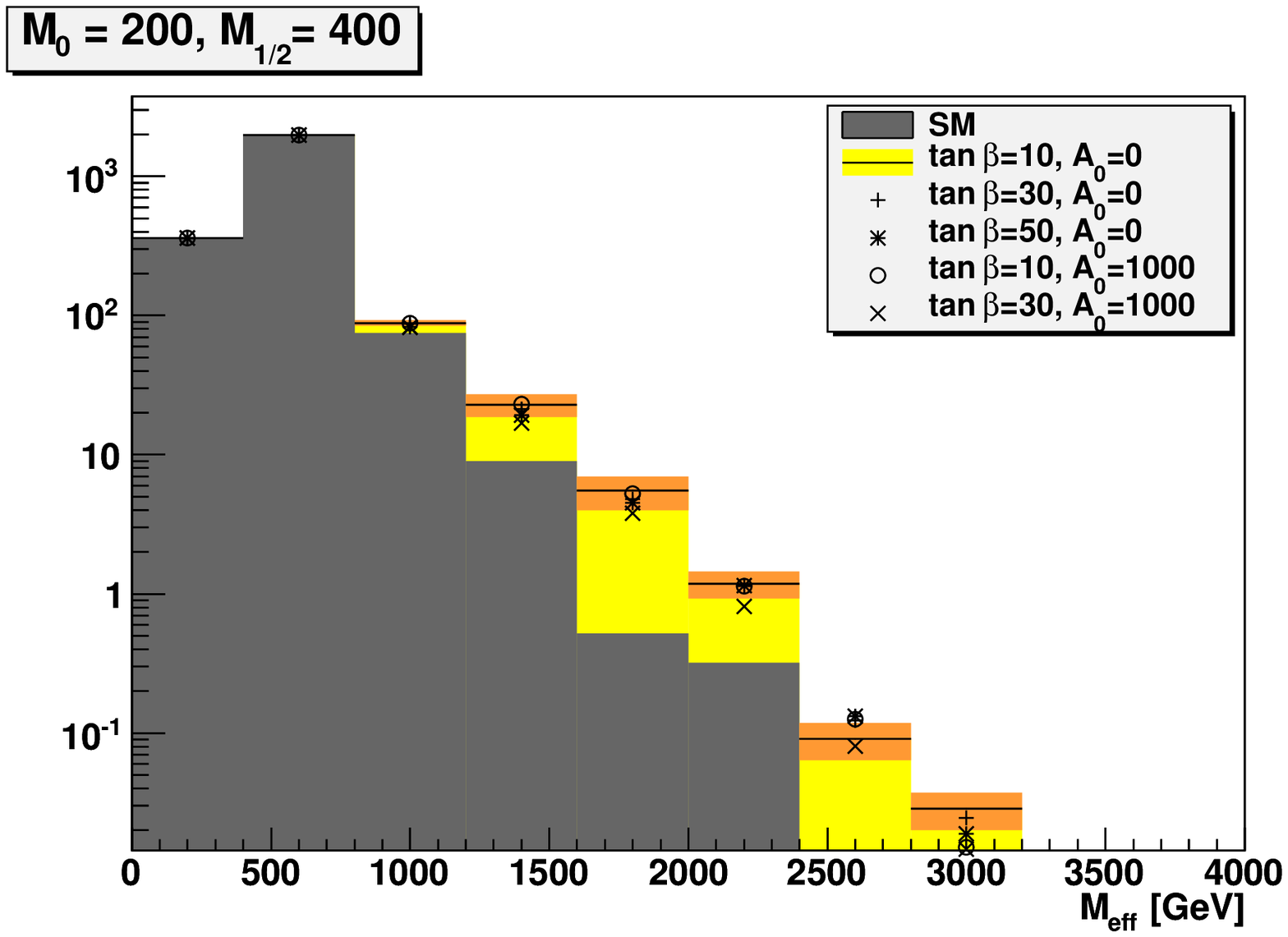}
    \label{fig:Syst_subfig2}
  }
  \caption{\sl Systematic check of the dependence of the simulated
    SUSY $M_{eff}$ spectrum on the parameters $A_0$ and $\tan\beta$
    for two different parameter points in $M_0$ and $M_{1/2}$. In
    \subref{fig:Syst_subfig1}, a point with low dependence on the
    parameters fixed in the grid are shown, showing a small variation
    of the predicted SUSY $M_{eff}$ spectrum well within
    systematics. In \subref{fig:Syst_subfig2}, a point with relatively
    large dependence is shown, which is still in agreement with the
    systematics. This shows the reliability of the application of the
    grid in $M_0, M_{1/2}$ in the fit.}
  \label{fig:Syst}
\end{figure}

On the signal, we apply a systematic uncertainty of 30\,\%, covering
both the uncertainty in the calculation of the cross section and the
remaining model dependence. The fit presented in
Section~\ref{sec:results} uses a grid spanned in $M_0$ and $M_{1/2}$
for the model prediction of the $M_{\rm eff}$ spectrum. In between the
model points, a bi-linear interpolation is used. The variation of the
$M_{\rm eff}$ spectrum with the remaining parameters $\tan\beta$ and
$A_0$ is shown in Fig.~\ref{fig:Syst}. The variations are clearly
compatible with the systematic uncertainty shown as the orange band.

\subsection{Statistics}\label{sec:statistics}

We use a likelihood ratio technique to calculate an expected
$CL_{s+b}$ for the non-observation of a signal at LHC. This confidence
level is then transferred into a contribution to the $\chi^2$ of the
frequentist fit. For a detailed description of the statistical method,
see Ref.\,\cite{Bechtle:2011dm}
 

\begin{figure}[t]
  \centering
  \subfigure[ ]{
    \includegraphics[width=0.45\textwidth]{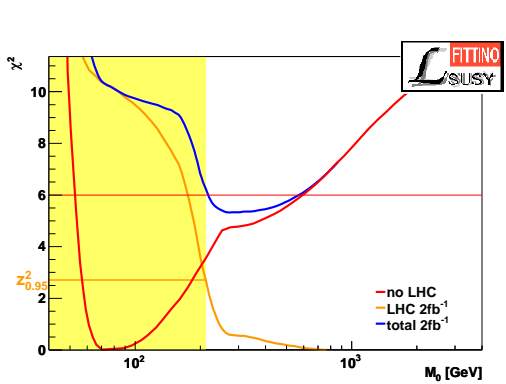}
    \label{fig:Agreement_subfig1}
  }
  \subfigure[ ]{
    \includegraphics[width=0.45\textwidth]{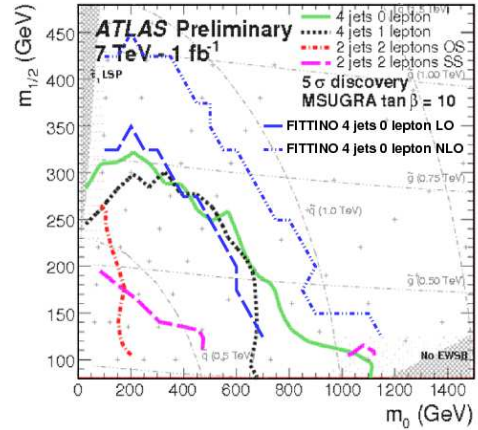}
    \label{fig:Agreement_subfig2}
  }
  \caption{\sl \subref{fig:Agreement_subfig1}: Interplay between the
    LHC and the pre-LHC contribution to the global
    $\chi^2$. \subref{fig:Agreement_subfig2}: Comparison between the
    discovery potential based on the ATLAS MC study (green line) and
    the LO result based on our implementation of the simulation of the
    detector and the analysis (dashed blue line).  }
  \label{fig:Agreement}
\end{figure}

As shown in Fig.~\subref{fig:Agreement_subfig1}, this technique
transfers the exact statistical power of the LHC search into a
contribution to the $\chi^2$. Thus, the global fit can find the exact
minimum and the exact uncertainties arising from the interplay between
the LHC contribution (orange) and the contribution from pre-LHC
observables (red). It can be seen that naturally LHC prefers high
SUSY mass scales, whereas the precision results prefer low scales, and
that the LHC contribution does not provide a considerably steeper
$\chi^2$ profile than the other results. The blue line represents the
combined $\chi^2$.

Very good agreement is achieved between the results
presented here and the official ATLAS study\,\cite{Atlas-pub}. Also,
the limit derived from our implementation agrees with the actual
search result for ${\cal L}^{int}=35\,\mathrm{pb}^{-1}$ of data within
a $1\,\sigma$ fluctuation of the background.

\section{Results}\label{sec:results}

The following results are obtained from global fits of the mSUGRA to
the observables described in Sec.~\ref{sec:model}. $\mathrm{sign}\mu=+1$ is
assumed for all fits due to the observed value of $(g-2)_{\mu}$. For a
more detailed analysis of the dependence of the pre-LHC-era fit on
$\mathrm{sign}\mu$, see Ref.\,\cite{Bechtle:2009ty}. For the LHC, integrated
luminosities ${\cal L}^{int}=0.035,1,2,7\,\mathrm{fb}^{-1}$ are
assumed, the first of those corresponding to the presently published
analyses, while the last corresponds to a reasonable expectation for
the available data set in 2011/2012.

\begin{figure}[t]
  \centering
    \includegraphics[width=0.5\textwidth]{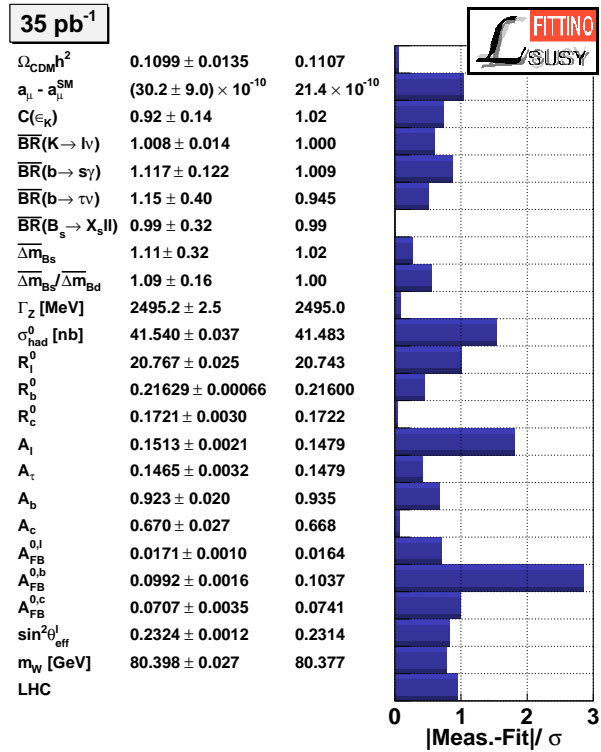}
  \caption{\sl All observables in the global fit and their pulls are
    shown 
    for ${\cal
      L}^{int}=35\,\mathrm{pb}^{-1}$. Excellent agreement is observed.}
  \label{fig:Pulls1}
\end{figure}

\begin{figure}[t]
  \centering
  \subfigure[ ]{
    \includegraphics[width=0.45\textwidth]{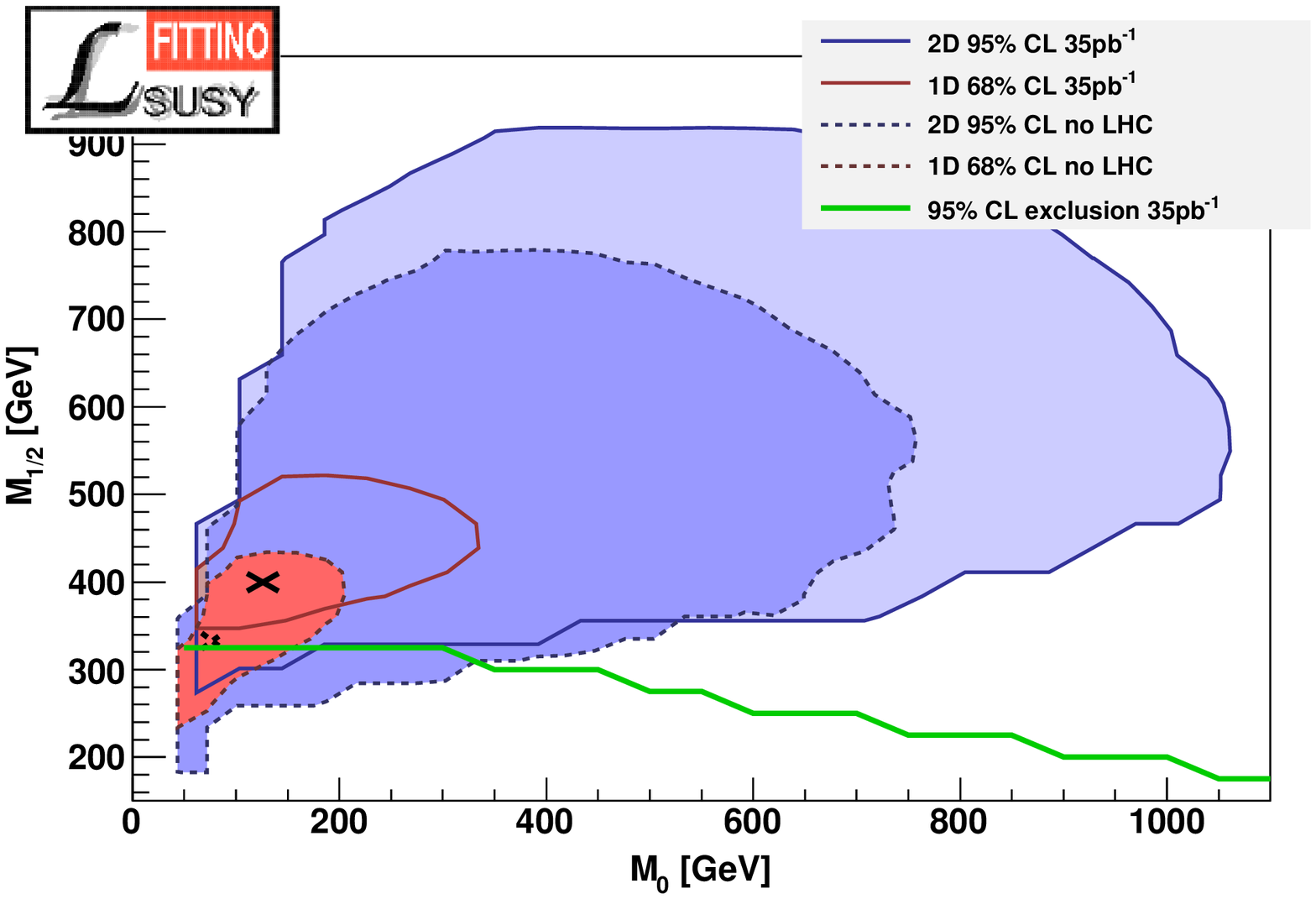}
    \label{fig:Results1_subfig1}
  }
  \subfigure[ ]{
    \includegraphics[width=0.45\textwidth]{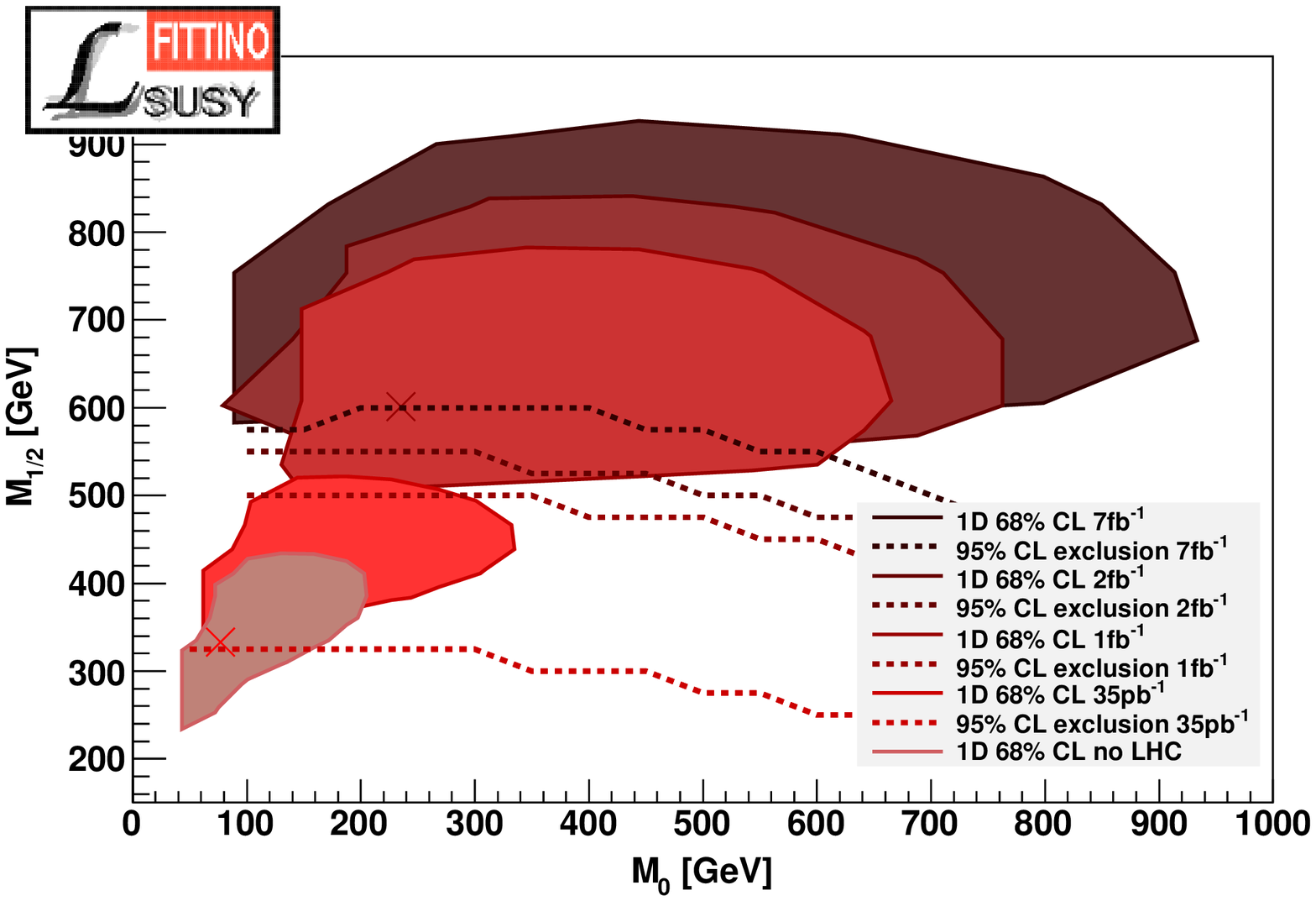}
    \label{fig:Results1_subfig2}
  }
  \caption{\sl The allowed parameter range for the fits without LHC
    and for the fit with our implementation present luminosity is
    shown in \subref{fig:Results1_subfig1}. The tension between the
    two fits is observed to be very moderate. In
    \subref{fig:Results1_subfig2}, the evolution of the
    $\Delta\chi^2=1$ area with increasing luminosity is shown. As
    expected, it moves to higher values of $M_0$ and $M_{1/2}$, and in
    addition the uncertainties grow very significantly.}
  \label{fig:Results1}
\end{figure}

For fits without LHC and for ${\cal L}^{int}=35\,\mathrm{pb}^{-1}$,
excellent agreement between the data and the mSUGRA model is
found. The pulls of the variables in the fit are shown in
Fig.~\ref{fig:Pulls1} for ${\cal L}^{int}=35\,\mathrm{pb}^{-1}$. More
importantly, there still is a significant agreement between the
resulting parameter ranges from the two fits, as shown in
Fig.~\ref{fig:Results1_subfig1}. While the LHC just excludes the best
fit point for the fit without LHC data, the $\Delta\chi^2=1$ areas do
still overlap significantly, and there is a large overlap in the
$\Delta\chi^2=5.99$ area, corresponding to a 95\,\%~CL in two dimensions. 

\begin{table}[t]
  \caption{\sl Overview of the best fit points for all considered LHC luminosities. 
    The values of $\chi^2/ndf$ underline that mSUGRA can not be excluded in the first 
    2 years of LHC.\vspace*{1ex}}\label{tab:chi2ndf}
  \begin{center}
    \begin{tabular}{|l|cccc|cc|}
      \hline
      ${\cal L}^{int}/\mathrm{fb}^{-1}$ & $M_0$ & $M_{1/2}$ & $\tan\beta$ & $A_0$ & $\chi^2/ndf$ & ${\cal P}-Value$ \\
      \hline
      0     & 77.1  & 332.8 & 12.8 & 426.2 & $18.9/20$ & $53.1\,\%$    \\
      0.035 & 125.9 & 399.8 & 17.3 & 742.3 & $20.4/21$ & $49.8\,\%$             \\
      1     & 235.1 & 601.0 & 31.1 & 626.8 & $23.7/21$ & $30.9\,\%$             \\
      2     & 254.1 & 647.1 & 30.2 & 770.7 & $24.2/21$ & $28.3\,\%$             \\
      7     & 402.7 & 744.1 & 43.1 & 780.7 & $25.0/21$ & $24.6\,\%$             \\           
      \hline
    \end{tabular}
  \end{center}
\end{table}

The fact that the current LHC analyses put little pressure on SUSY is
also evident from Tab.~\ref{tab:chi2ndf}, which shows the best fit
points of the five fits together with the observed $\chi^2/ndf$ values
and the corresponding ${\cal P}$-values (the latter being only for
completeness, since it is technically not proven that the expected fit
results follow a $\chi^2$ curve, due to significant non-linearities
both in the LHC limits and in the relation between parameters and
observables). In any case, the change in $\chi^2/ndf$ is very moderate
when going from the fit without LHC to the fit with ${\cal
  L}^{int}=35\,\mathrm{pb}^{-1}$.

This observation contradicts the disappointment about the
non-observation of SUSY at LHC with ${\cal
  L}^{int}=35\,\mathrm{pb}^{-1}$, which is mostly based on finetuning
arguments or Bayesian discussions of the size of the available
parameter space for arbitrary priors. Without those more subjective
measures of the attractiveness of a theory, even the highly constrained
mSUGRA is still in natural agreement with the data. To the contrary,
squark masses of around $1$~TeV or slightly higher are a welcome
ingredient to lift the mass of the lightest, SM-like Higgs boson above
the LEP limit.

For higher assumed LHC luminosities, still assuming no observation of
new physics, the $\Delta\chi^2=1$ areas of the fits do start to
deviate significantly from each other, as evident from
Fig.~\ref{fig:Results1_subfig2}. This corresponds to a building
tension in the fit between mainly $(g-2)_{\mu}$ and $\Omega_{\chi}$,
pushing mSUGRA to lower scales via the gaugino and slepton sector, and
the LHC, pushing mSUGRA to higher scales via the more direct limit on
the squark and gluino mass scale. This results in a degradation of
$\chi^2/ndf$, as evident from Tab.~\ref{tab:chi2ndf}. However, even
for ${\cal L}^{int}=7\,\mathrm{fb}^{-1}$, mSUGRA can not be excluded
with the given observable set and SUSY searches alone. This tension is
expected to be significantly weaker for more general SUSY models.

\begin{figure}[t]
  \centering
  \subfigure[ ]{
    \includegraphics[width=0.45\textwidth]{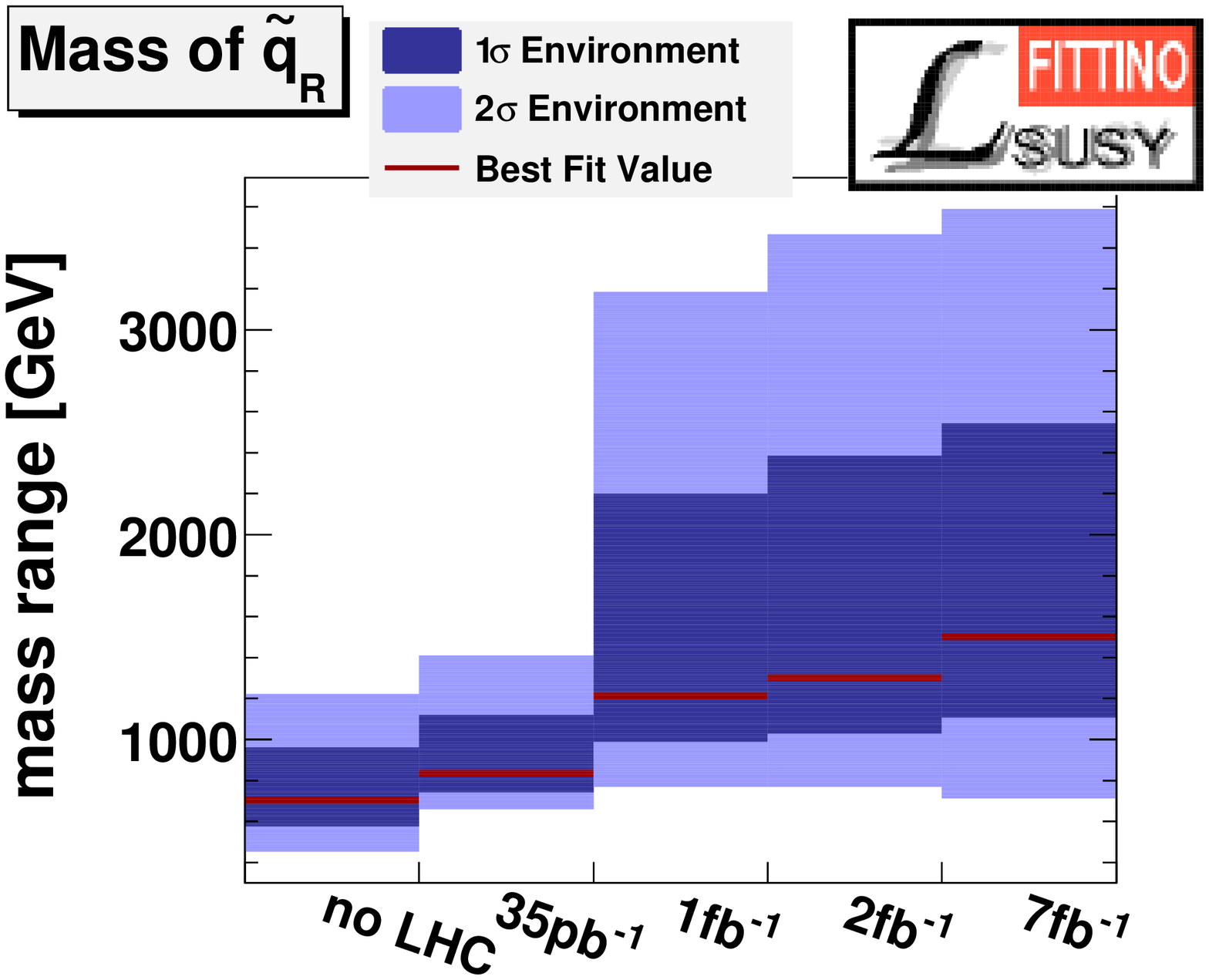}
    \label{fig:Results2_subfig1}
  }
  \subfigure[ ]{
    \includegraphics[width=0.45\textwidth]{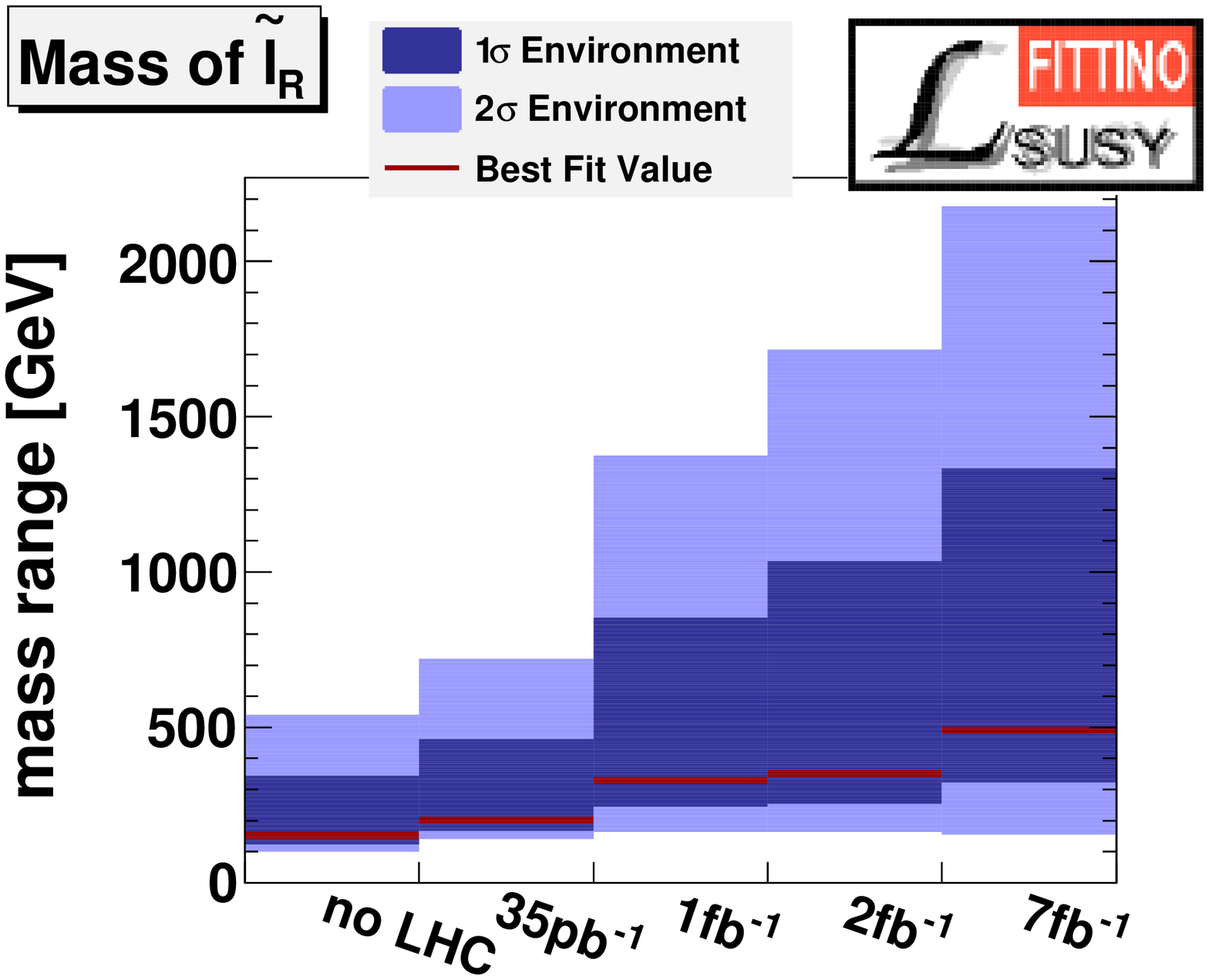}
    \label{fig:Results2_subfig2}
  }
  \caption{\sl For the different LHC luminosities, this figure shows
    the allowed mass ranges of squarks in
    \subref{fig:Results2_subfig1} and of sleptons in
    \subref{fig:Results2_subfig2}. While the former is quite model
    independent, the latter strongly depends on assumptions in the
    mSUGRA model.}
  \label{fig:Results2}
\end{figure}

Even though the tension is rising, SUSY cannot be excluded at the LHC
in the first two years of running. The interesting observation here is
that the inclusion of the LHC exclusion to the fit only has a very
moderate effect on the lower mass bound of the sparticles, as shown in
Fig.~\ref{fig:Results2}. The by far non-trivial result from the fit,
however, is the fact that the upper bound on the sparticle masses
depends very strongly on including the LHC into the fit. The reason
for this behavior can be seen in
Fig.~\ref{fig:Agreement_subfig1}. The $\chi^2$ surface is influenced
by LHC only for $M_{0}<1.5$~GeV, it remains independent of the LHC
luminosity above that value. However, there it is significantly more
flat than close to the minimum of the fit without LHC. Including the
LHC cuts away the low $\chi^2$ values, shifting up the
$\Delta\chi^2=4$ area significantly into shallower areas of the
$\chi^2$ profile. Therefore, non-trivially, the upper mass bounds on
the sparticles increase very strongly, allowing mSUGRA to escape the
LHC detection to higher mass regions.

\begin{figure}[t]
  \centering
  \subfigure[ ]{
    \includegraphics[width=0.45\textwidth]{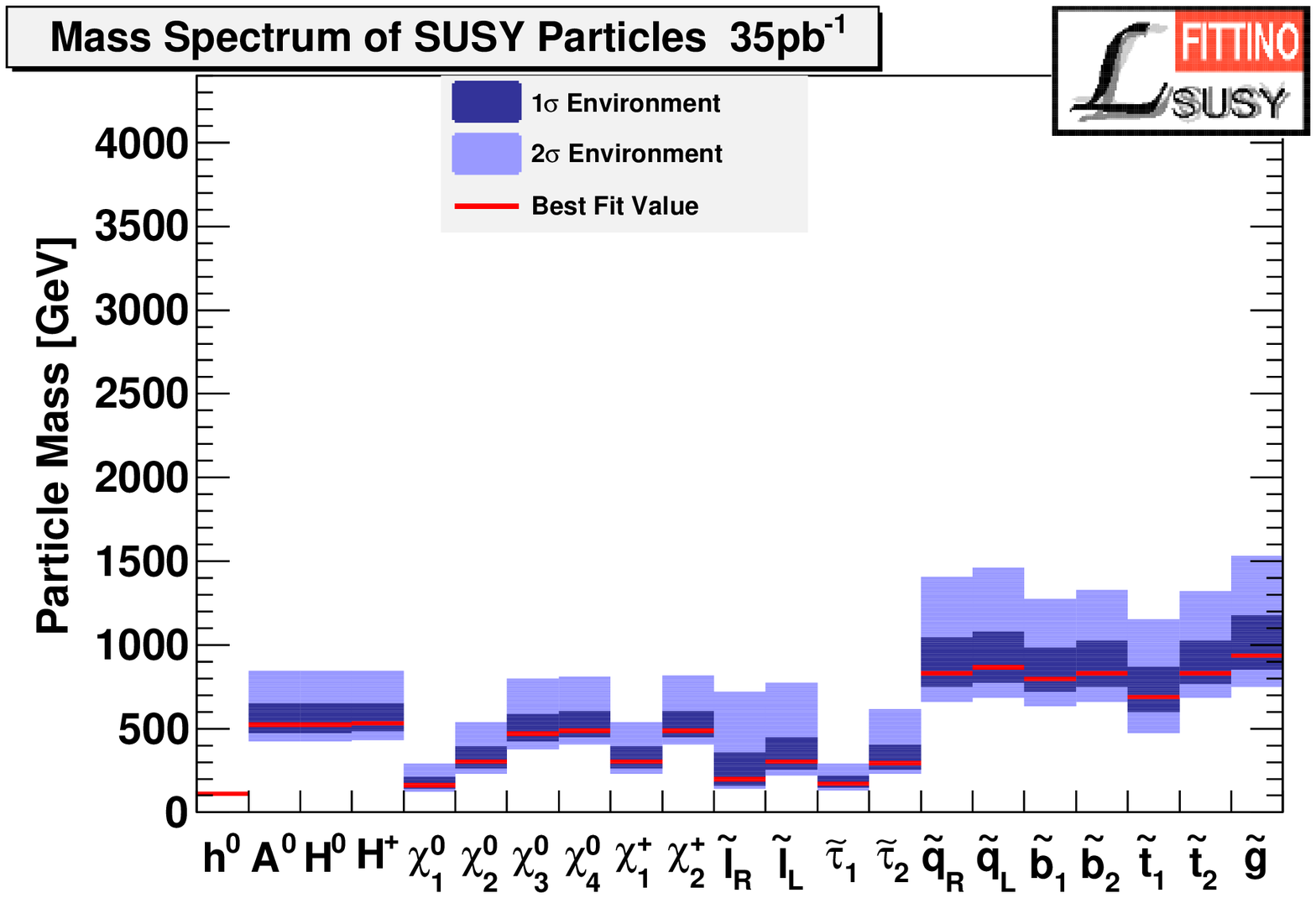}
    \label{fig:Results3_subfig1}
  }
  \subfigure[ ]{
    \includegraphics[width=0.45\textwidth]{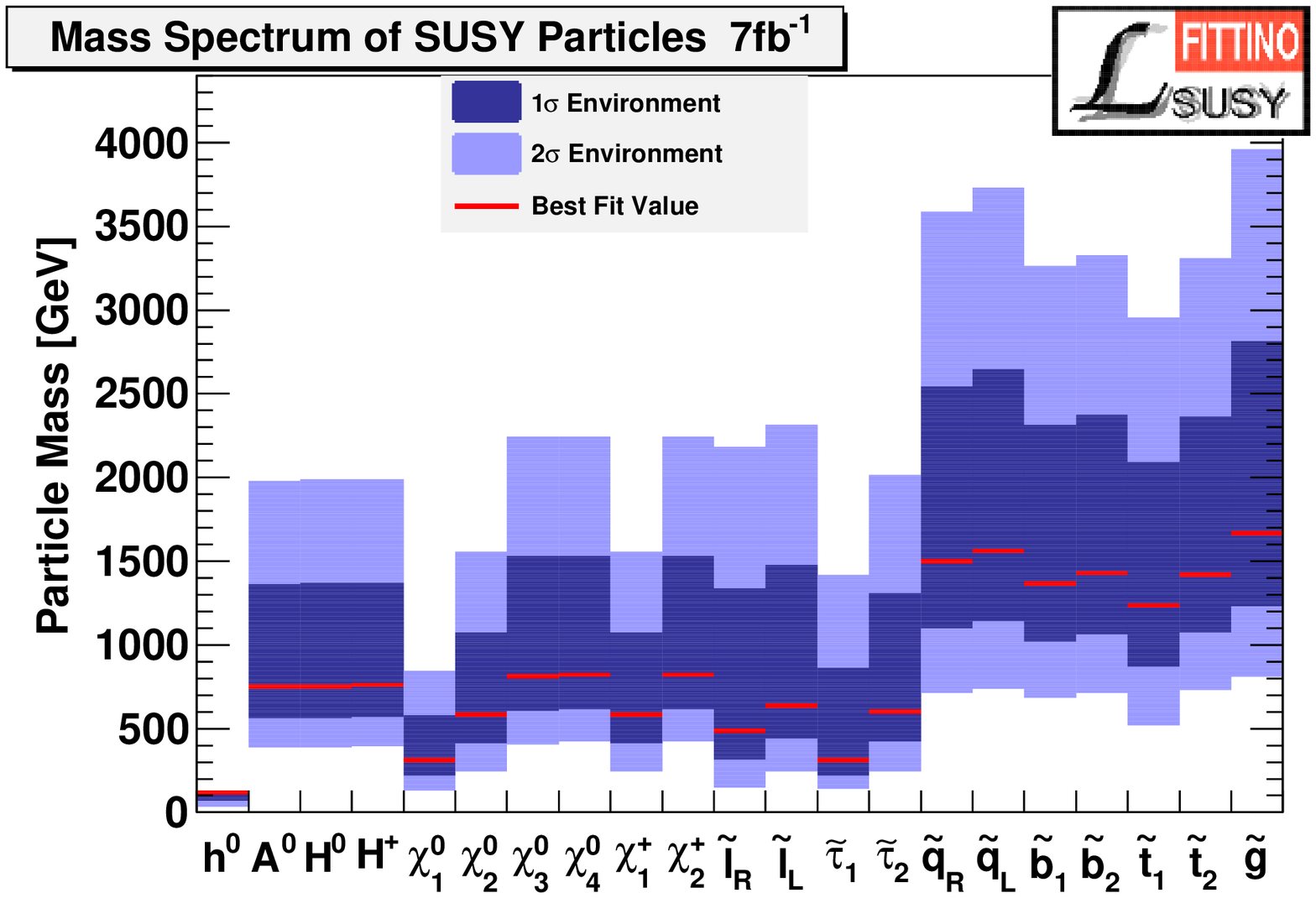}
    \label{fig:Results3_subfig2}
  }
  \caption{\sl Comparison of the allowed mass ranges of all sparticles
    and Higgs bosons for the fits with ${\cal
      L}^{int}=35\,\mathrm{pb}^{-1}$ in \subref{fig:Results3_subfig1}
    and  ${\cal
      L}^{int}=7\,\mathrm{fb}^{-1}$ in \subref{fig:Results3_subfig2}.}
  \label{fig:Results3}
\end{figure}

Fig.~\ref{fig:Results3} shows the same for all sparticles and Higgs
bosons, but only for ${\cal L}^{int}=35\,\mathrm{pb}^{-1}$ and ${\cal
  L}^{int}=7\,\mathrm{fb}^{-1}$. The interesting observation is that
the only particle of which the allowed mass range does not change is
the SM-like Higgs boson $h^0$, which is bound in mSUGRA at
$m_{h^0}<135$~GeV. Therefore, the only chance for an exclusion of
mSUGRA and many other SUSY breaking scenarios can be obtained via
SM-like Higgs searches at the Tevatron and LHC.

\begin{figure}[t]
  \centering
  \subfigure[ ]{
    \includegraphics[width=0.45\textwidth]{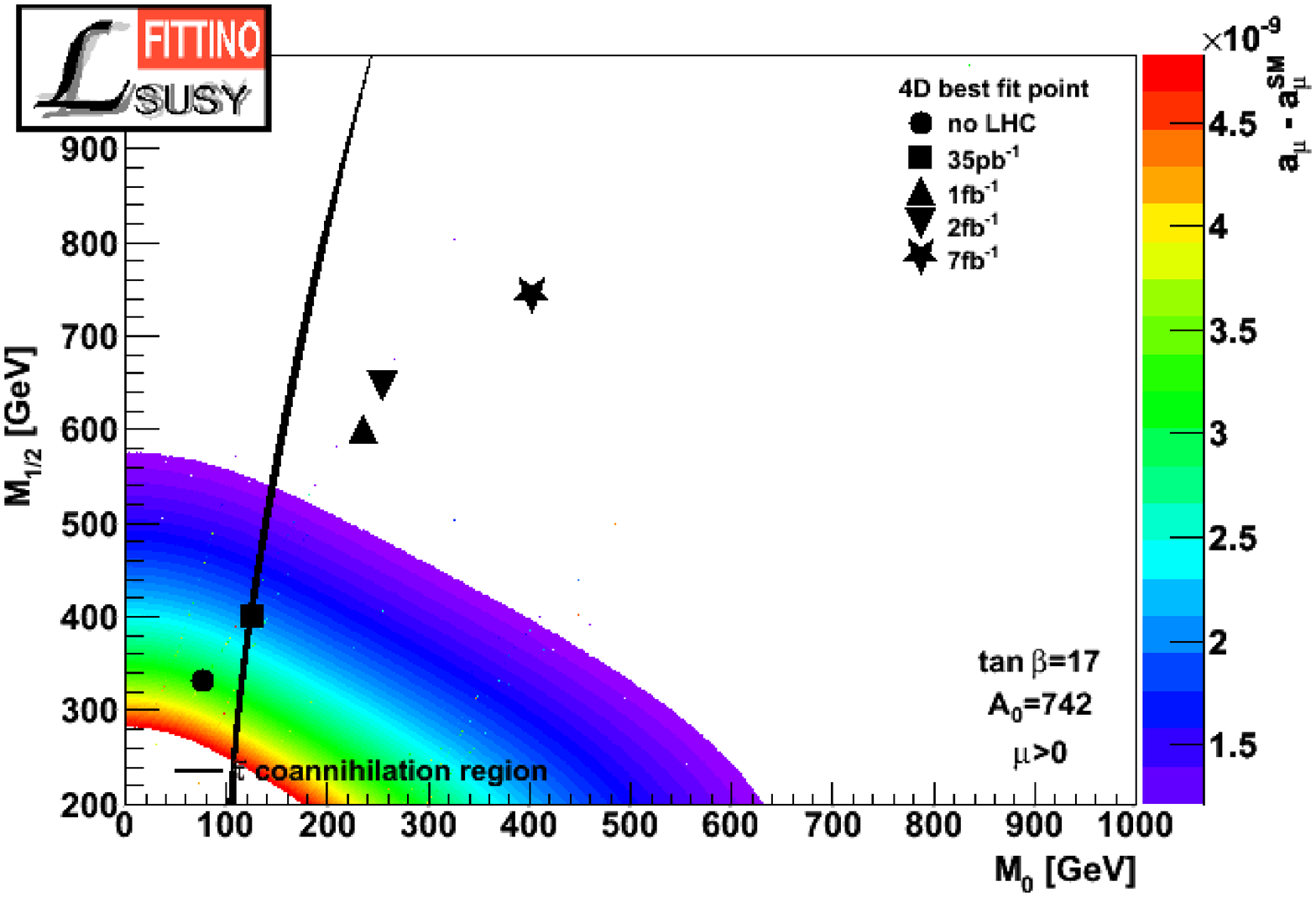}
    \label{fig:Results4_subfig1}
  }
  \subfigure[ ]{
    \includegraphics[width=0.45\textwidth]{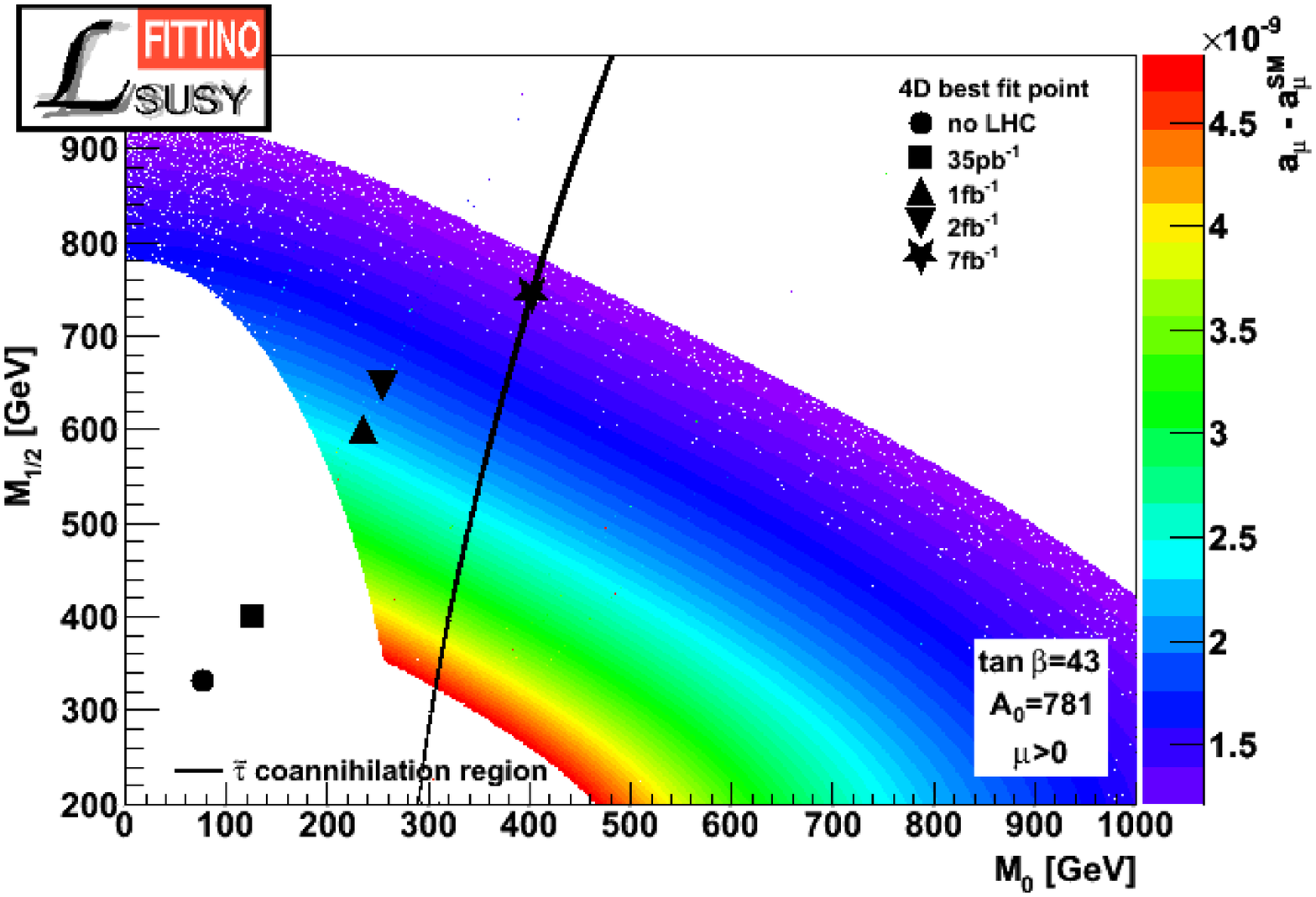}
    \label{fig:Results4_subfig2}
  }
  \caption{\sl Explanation, of how the interplay of $(g-2)_{\mu}$,
    $\Omega_{\chi}$ and the LHC moves the best fit point of
    $\tan\beta$ to significantly higher values for higher LHC
    luminosities. In \subref{fig:Results4_subfig1}, the situation for
    $\tan\beta=17$ is shown, where $(g-2)_{\mu}$ and $\Omega_{\chi}$
    are in good agreement with the data for low $M_0$ and
    $M_{1/2}$. If LHC should exclude those low regions of $M_0$ and
    $M_{1/2}$, a higher value of $\tan\beta$ is necessary to reconcile
    $(g-2)_{\mu}$ and $\Omega_{\chi}$ with the data, as shown in
    \subref{fig:Results4_subfig2} for $\tan\beta=43$.}
  \label{fig:Results4}
\end{figure}

One interesting observation is the fact that the LHC pushes the best
fit point of $\tan\beta$ to significantly higher values than observed
for the fit without LHC. This is interesting since it is shown in
Fig.~\ref{fig:Syst} that the LHC limit in the chosen search channel
does not depend significantly on $\tan\beta$. The increase however is
an interesting showcase of an interplay between low-energy precision
observables, cosmological observables and direct limits from the
LHC. This is described in Fig.~\ref{fig:Results4}, showing the
co-annihilation region which is mainly responsible for a good fit of
$\Omega_{\chi}$ (another region with some contribution from the Higgs
funnel also is allowed at very large $\tan\beta$) and the predicted
values of $(g-2)_{\mu}$ for $M_{0}$ and $M_{1/2}$. In
Fig.~\ref{fig:Results4_subfig1}, $\tan\beta=17$ is used and the two
observables agree with the measurements for low mass scales. In
Fig.~\ref{fig:Results4_subfig2} and $\tan\beta=43$, the low mass
scales can be excluded by the LHC, retaining agreement with
$\Omega_{\chi}$ and $(g-2)_{\mu}$ at high mass scales. Thus, the
exclusion of low mass scales pushes mSUGRA to higher values of
$\tan\beta$, since only then the pre-LHC-era observables can be
correctly described. This is an interesting observation, since also in
the detailed study of theoretical uncertainties of mSUGRA models up to
now the focus was on the low to intermediate $\tan\beta$ region. A
further non-observation of SUSY at the LHC would highlight the
importance of understanding SUSY precision calculations at high values
of $\tan\beta$.


\section{Conclusions}\label{sec:conclusions}

We have presented a global mSUGRA analysis of supersymmetric models
which includes the current low-energy precision measurements, the dark
matter relic density as well as potential LHC exclusion limits from
direct SUSY searches in the zero-lepton plus jets and missing
transverse energy channel.  

We conclude that non-trivially it is possible to reconcile the
supersymmetric description of low-energy observables and the dark
matter relic density with a non-observation of supersymmetry in the
first phase of the LHC with acceptable $\chi^2/ndf$ values, despite
some tension building up in a combined fit within mSUGRA.

While our study is exploratory in the sense that it is based on one
search channel only, and on a simplified description of the LHC
detectors, it clearly demonstrates the potential of the first phase of
LHC running at 7\,TeV in 2011/12 to constrain supersymmetric models
and the sparticle mass spectrum, or to discover such models.

However, the interesting fact that including LHC limits in the global
fit significantly increases the upper bounds on the sparticle masses
make it impossible to exclude mSUGRA in the first two years at LHC
based on SUSY searches. Excluding the model could however be possible
using Higgs boson searches.

\section*{Acknowledgments}

\noindent We thank Sascha Caron and Werner Porod for valuable
discussions. This work has been supported in part by the Helmholtz
Alliance ``Physics at the Terascale'', the DFG SFB/TR9 ``Computational
Particle Physics'', the DFG SFB 676 ``Particles, Strings and the Early
Universe'', the European Community's Marie-Curie Research Training
Network under contract MRTN-CT-2006-035505 ``Tools and Precision
Calculations for Physics Discoveries at Colliders'' and the Helmholtz
Young Investigator Grant VH-NG-303. MK thanks the CERN TH unit for
hospitality.


\end{document}